\documentclass[%
 aip,
 amsmath,amssymb,
reprint,%
tikz
]{revtex4-2}

\usepackage{graphicx,subcaption}%
\usepackage{dcolumn}%
\usepackage{bm}%

\usepackage[utf8]{inputenc}
\usepackage[T1]{fontenc}
\usepackage{mathptmx}

\begin{document}

\preprint{AIP/123-QED}

\title[High Speed Flows with PonD: Boundary Conditions]{High Speed Flows with Particles on Demand: Boundary Conditions}

\author{Abhimanyu Bhadauria}
\author{Ilya Karlin}
\affiliation{Department of Mechanical and Process Engineering, ETH Zurich, 8092 Zurich, Switzerland%
}

\date{\today}%

\begin{abstract}
The particles on demand (PonD) method is a new kinetic theory model that allows for simulation of high speed compressible flows. While standard Lattice-Boltzmann is limited by a fixed reference frame, significantly reducing the range of applicable of Mach numbers, PonD takes advantage of adaptive reference frames to get rid of the restrictions of standard LB and is able to simulate flows at high speeds and with large temperature gradients. 
Previously, PonD has been shown to be a viable alternative for simulation of flows with strong discontinuities and for detonation modelling.  
However, treatment of flows with complex boundaries has been lacking. Here, we present PonD augmented with a non-equilibrium extrapolation based boundary condition. We present several compressible test cases such as shock-vortex interaction in the Schardin's Problem and supersonic flow over a two-dimensional cylinder at Mach numbers up to 5. We observe that the results agree well with literature, paving the way for a kinetic theory based approach for simulating compressible flows in realistic scenarios.
\end{abstract}

\maketitle

\section{\label{sec:level1} Introduction}

All fluid are compressible to a certain extent, however, at low Mach numbers $M \leq 0.3$, the effects of compressibility can be largely ignored, and as such flow can be treated as being incompressible without any loss of accuracy. As the Mach number increases, however, modelling the effect of compressibility becomes important. The modelling of high-speed compressible flows are especially relevant to the design of aircraft, rockets, re-entry vehicles etc. Modelling of reactive flows inside internal combustion engines or jet engines also require an accurate numerical model to simulate compressible flow. 
Besides application in aerospace engineering, the flow of blood inside our bodies is also an example of compressible flow. The importance of accurate modelling of compressible flows therefore cannot be overstated. 
Traditionally such flows have been modelled with classical computational fluid dynamics methods such as the finite volume or finite element.

Recently however, the lattice Boltzmann method has emerged as an alternative to classical  computational fluid dynamics (CFD) methods. 
Building upon the discrete form of the Boltzmann transport equation, the lattice Boltzmann method (LBM) uses a bottom-up approach to arrive at the Navier-Stokes (NS) equation for fluid dynamics.
LBM has been gaining prominence in the domain of CFD, with promising results having been shown in a wide array of flows ranging from turbulence \cite{CHIKATAMARLA2010,FRAPOLLI20182}, multiphase flows\cite{A.Mazloomi2015} and multi-component flows \cite{Sawant2020} to rarefied gas flows \cite{Staso2016} to name a few.

However, the applications of LBM remained restricted to low speed incompressible flows. 
At high speeds or in case of large temperature deviations (as is common in compressible flows), the standard lattice (D2Q9 in 2D and D3Q27 in 3D) {suffers from insufficient isotropy} which induce errors in the fluid stress tensor and breaks Galilean invariance. {Mathematically, the increased deviation from lattice temperature and the rest velocity leads to an error in the Chapman-Enskog expansion}, which means that the discrete Boltzmann equation does not derive NS equations at the macroscopic scale. This severely limits high speeds and temperature deviations in the fluid. 

Several solutions have been offered to tackle this problem and make LB suitable for compressible flows. 
A straightforward solution to solve the problem of insufficient isotropy is to increase the number of lattice velocities, conserving more moments \cite{Chikatamarla2006,XIAOWEN2006,Chikatamarla2009,Siebert2008,CHIKATAMARLA2010}. While this does work, it comes at the cost of an increasingly prohibitive computational complexity.

Another solution that works for some flows is to introduce a shift in the lattice velocities \cite{Frapolli2016Shift,Hosseini_2019Shift}, thereby shifting the error to be centered around the free stream velocity. This method can be used to model arbitrarily high mach number flows but it is important to remember that while it reduces the deviations in the higher order moments around the free-stream (shifted) velocity, errors do accumulate in regions far from the shift velocity (eg. walls). Therefore it does not increase the acceptable velocity range by a large amount. 

The two population model was introduced with the aim of tackling incompressible flows with temperature variation. The temperature dynamics were uncoupled by introducing a second populations for conservation of either total energy or internal energy \cite{HE1998282,KARLIN2013}. 

To fully maximise the velocity range, correction terms were introduced, to cancel out the errors in the higher order moments that arise from velocity and temperature deviation. Several authors have formulated and recommended such correction terms \cite{Feng_2015,Huang_2019}. Saadat et. al \cite{Saadat2019} extended the consistent two population thermal model \cite{KARLIN2013} to compressible flows using simplified correction terms. The effect of deviation from reference temperature and correction terms to restore Gallilean invariance was studied by Hosseini et al. \cite{Hosseini_2020}

The use of correction terms in the two population setting \cite{Saadat2020,saadat2021extendedGas} made it possible to simulate transonic and low-supersonic flows while still using the D2Q9 lattice. This methods was also extended with moving boundaries \cite{Saadat2020a,bhadauria2021} and showed promising results up to $M\sim 2$.
However, with all these improvements, the maximum Mach number that could be achieved was still limited to $M\sim 2$.

The recently introduced Particles on Demand (PonD) method by Dorschner et al. \cite{Dorschner2018b} tackled both the velocity and temperature variation limits that still plagued the previous schemes. In the previously mentioned schemes with shifted stencils \cite{Frapolli2016Shift}, a uniform shift was applied to the entire domain, leading to a fixed reference frame, which meant that the errors were centered around the shift velocity and lead to instability in regions which had a velocity away from the shift ( such as no slip walls ). Moreover, the problem of temperature variation was never adequately tackled. PonD seeks to address both these issues in LB. In PonD, the lattice velocities at each node are shifted by the local velocity, and scaled by the local temperature, which leads to an adaptive co-moving reference frame. 
The result in a scheme that derives the NS equations exactly and is Galilean invariant, at all speeds and temperature ratios. 
Once the restrictions on velocity and temperature are lifted, PonD is able to simulate high speed compressible flows with large temperature and velocity gradients. 

Since its initial development, PonD has been implemented and validated for a wide variety of applications from  compressible multiphase flows with non-ideal fluids\cite{Reyhanian2020,Reyhanian2021,ReyhanianPHD}, to flows with strong shocks \cite{kallikounis2022particles}, to detonation \cite{sawantPonD_2022}. 
However, the focus has not yet shifted to boundary conditions for PonD.

In nature and engineering fluid flows that encounter solid obstacles are ubiquitous and therefore any model for simulating realistic fluid flows must be able to accurately resolve the effect that these obstacles have on to the flow. 
The canonical case of flow over a spherical body leading to a von Karman vortex street is well studied both experimentally and numerically. 
In aerospace engineering, the study of the flow over an aerofoil is an essential part of the design process of any aircraft. 
At higher speeds, such as for supersonic aircraft and hypersonic re-entry vehicles, the effect of heat transfer to the body becomes significant as well. 
Wall bounded compressible flows with combustion and detonation are also an intrinsic part of the design of gas turbine and rocket engines.
For obtaining a complete picture of a complex dynamic system, it is often necessary to model the coupling between the fluid and the solids as well.

Within the context of LB, several boundary conditions been proposed and implemented to extend its application to flows with complex boundaries. 

One of the simplest yet robust boundary condition in LB is the bounce-back scheme. While standard bounce-back is a conservative scheme and easily extended to second order accuracy it it's application is limited to flat boundaries.  

 Bouzidi et al. \cite{Bouzidi2001} proposed an extension to curved walls using bounce-back with linear and quadratic interpolation that maintains second order accuracy. The interpolated bounce-back was improved by several authors  \cite{Ginzburg2003,LALLEMAND2003406,Yu2012} by improving accuracy and extending it to moving bodies. The simplicity and robustness of bounce-back stems from the symmetry of the lattice and a simple reflection rule to obtain missing populations at the boundary.  It should be noted however that interpolated bounce-back loses its conservative nature due to interpolation.

A different approach to treating curved boundaries in LB was provided by Guo et al. \cite{Guo2002} by splitting the distributions on boundary nodes into their equilibrium and non-equilibrium parts. While the equilibrium is obtained using interpolated values for density and velocity at the wall, the non-equilibrium distribution is obtained via extrapolation from fluid neighbours. The scheme is 2nd order accurate, stable, and does not depend on a bounce-back like step to impose boundary conditions.

{ While the aforementioned boundary conditions are general and translate well to standard LB models quite easily, their extension to PonD presents certain challenges. Due to the aforementioned reference frames in PonD being completely adaptive and transient, there is an inherent asymmetry of lattice links, in both space and time. Since a majority of the LB boundary conditions rely on direct propagation along links, PonD requires some adaptation to be made in the boundary conditions. In this work, we propose one such method for implementing boundary conditions in PonD, that is based on the non-equilibrium extrapolation scheme of Guo et al.}

In section \ref{sec:PondModel} we give a brief introduction of PonD. The numerical implementation of the model including the interpolation schemes and reconstruction of the populations is described in detail in section \ref{sec:PondNumerics}. In section \ref{sec:BoundaryCondtions} we go over the details of the boundary conditions. We then present results in section \ref{sec:results} and the conclusion and outlook is drawn in section \ref{sec:conclusions}.

\section{\label{sec:PondModel} Kinetic Model: Particles on Demand }

We start by describing our kinetic model for simulating hydrodynamics, the particle on demand method (PonD). 
First, we define $\bm{u}$ and ${T}$ as the local velocity and temperature, respectively at each lattice node. We then define a frame of reference $\lambda$ as a pair of the local velocity and temperature, $\lambda = \{\bm{u},{T} \}$. 

The key difference between LB and PonD is that in PonD lattice discrete velocities are adaptive, scaled by the ratio of the root of the local temperature and the lattice temperature, and then subsequently shifted by the local fluid velocity at that node.  
For a standard LB lattice, in D dimensional space and with Q discrete velocities, we define the particle velocity $\bm{v}_i$, distinct from discrete velocity  $\bm{c}_i$,  at each node as:
\begin{equation}
    \label{eq:particleVel}
 \bm{v}_i = \sqrt\frac{T}{T_0}  \bm{c}_i + \bm{u},\ {i} = \{0,\dots,Q-1\}. 
\end{equation}

The local frame of reference $\lambda = \{ \bm{u},{T} \} $ is the co-moving reference frame which is the frame of interest in PonD, while in standard LB, the reference frame is the "rest" reference frame everywhere i.e $\bm{u} = 0 $ and ${T} = T_0$, where $T_0$ is the lattice temperature, a characteristic of the chosen lattice. In this work, we use the D2Q16 lattice for all simulations.
\begin{figure}
    \centering\includegraphics[width=\linewidth]{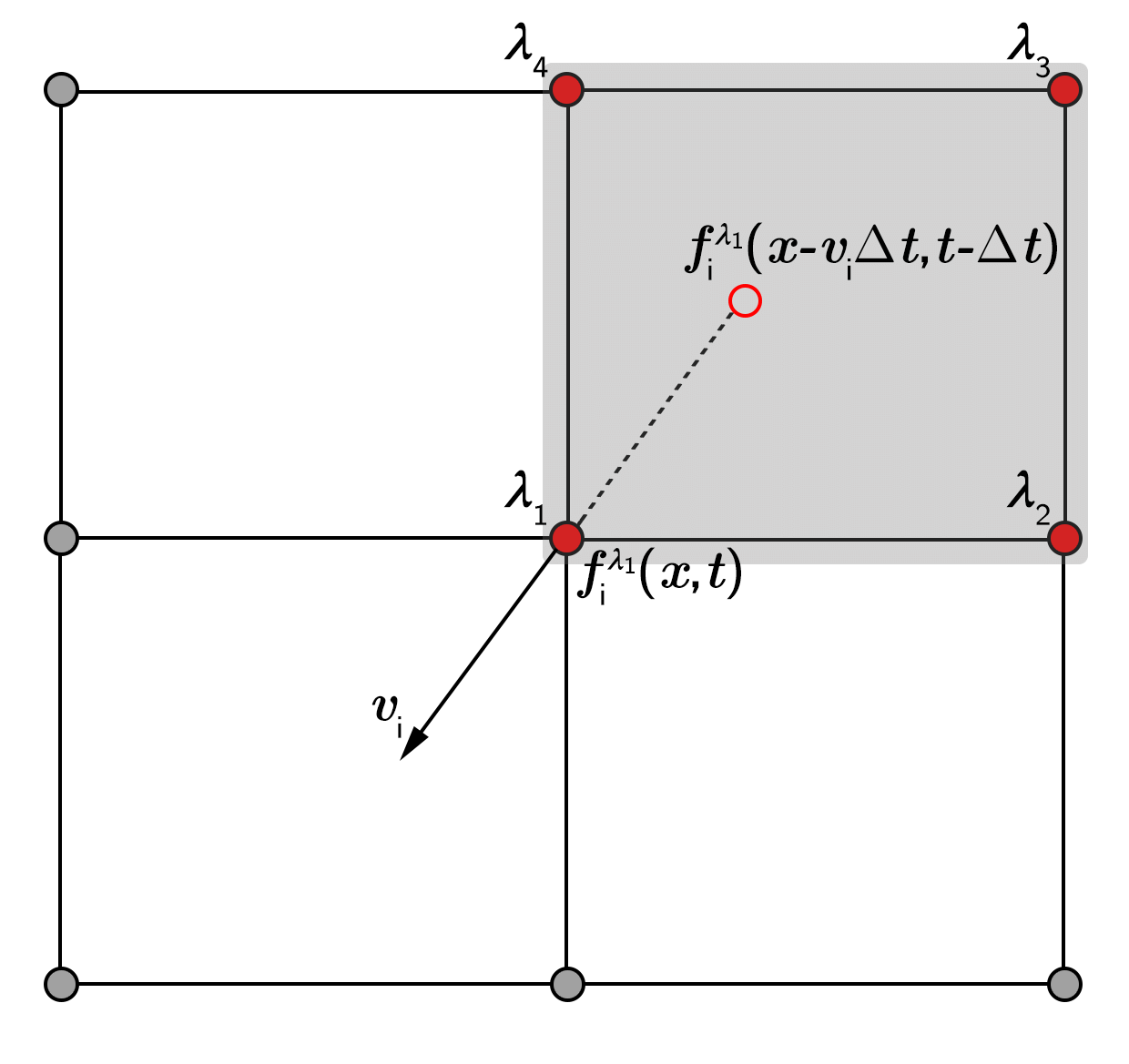}
    \caption{Semi Lagrangian Advection in PonD. Interpolation at the departure point (open circle) using the stencil (filled circles). In PonD, populations must be transformed to the destination frame before interpolation.  }
    \label{fig:pond_semiLag}
\end{figure}

Advection now occurs along the co-moving particle velocities which are fully adaptive and therefore no longer space filling. Similar to other off-lattice LB methods, the departure point, $\bm{x}_{\rm dp}$ may no longer land at a lattice node, therefore, advection needs to proceed in a semi-Lagrangian fashion. 

Populations at each node are defined at the local reference frame, and therefore care has to be taken during interpolation to ensure that the populations have first been transformed to a common reference frame before collision. 

In the next section we go over the details of the semi-Lagrangian advection, frame transformations, and collision in detail. 

\section{\label{sec:PondNumerics} Particles on Demand : Numerical Implementation }

\subsection{Collision}

In this work we employ the two-population model previously explored in Ref. \onlinecite{KARLIN2013,Saadat2019} in conjunction with PonD. 
The ${f}$-populations conserve mass and momentum, while the ${g}$-populations are used to conserve the internal energy. This allows for a variable Prandtl number, but in this work we limit ourselves to $\mathrm{Pr}=1$. 
\begin{eqnarray}
    \label{eq:f_collision}   f_i (\bm x, t) &= f_i (\bm x,t) + 2 \beta (f_i^{\rm eq}(\bm x,t) - f_i(\bm x,t)), \\
    \label{eq:g_collision}   g_i (\bm x, t) &= g_i (\bm x,t) + 2 \beta (g_i^{\rm eq}(\bm x,t) - g_i(\bm x,t))  
    \label{eq:collision}.
\end{eqnarray}

In the co-moving reference frame, equilibrium populations ${f_i^{\rm eq}}$ and ${g_i^{\rm eq}}$ are exact and do not depend on the velocity. 
The complete derivation of the Chapman-Enskog expansion can be found in Ref. \onlinecite{Dorschner2018b}. 
\begin{align}
    &f_i^{\rm eq}(\bm{x},t) = w_i\rho(\bm{x},t), \\
    &g_i^{\rm eq}(\bm{x},t) = (2 C_v - D)  T(\bm{x},t) f_i^{\rm eq}(\bm{x},t).  
\label{eq:Exactequilibrium}
\end{align}
Here, $C_V$ is the specific heat at constant volume and D is the dimension. 
The constant weights $w_i(\bm{c}_i)$ are characteristic of the chosen lattice, and $\beta$ is the relaxation parameter defined as : 
\begin{align}
    \beta=\frac{T \Delta t}{2 \nu + T \Delta t },
\label{eq:betaRelax}
\end{align}
where $\nu$ is the kinematic viscosity. 

The standard conservation laws apply here,
\begin{align}
    \rho&=\sum_{i=0}^{Q-1} f_i, \\
    \rho \bm{u}&=\sum_{i=0}^{Q-1} f_i \bm{v}_{i},\\
    2 \rho E &= 2 \rho \left(C_v T + \frac{{u}^2}{2}\right) = \sum_{i=0}^{Q-1} g_i + \sum_{i=0}^{Q-1} f_i {v}^2_{i}.
\label{eq:conservationLaws}
\end{align}
One must note that the populations here should all be in the same reference frame. 

\subsection{Semi-Lagrangian Advection}

For a general point advection, we write ${\bm{x}_{\rm dp}}$ as the departure point of a population at a lattice node at location $\bm{x}$,  with particle velocity $\bm{v}_i$ : 
\begin{equation}
     \bm{x}_{\rm dp}= \bm{x} - \bm{v}_i {dt}.
\end{equation}

Then, the i-th population $f_i^\lambda$ at position $\bm{x}$ and time $t$ 
\begin{equation}
\centering  f_i^\lambda( \bm{x}, t) = f_i^\lambda( \bm{x}_{\rm dp}, t-dt), \hspace{0.2cm} {i} = \{0,Q-1\}. 
\end{equation}
Here the superscript $\lambda$ indicates that the populations are computed at the reference frame, $\lambda$.

The particles velocities in PonD, as defined in equation \ref{eq:particleVel}, are adapted to the local flow velocity and temperature, and hence exact advection cannot be guaranteed. Therefore, $f_i^\lambda( \bm{x}_{\rm dp}, t-dt)$ has to be interpolated using known populations at neighbouring lattice nodes. 

 In general we can write for a $p$-sized interpolation kernel $W(\bm{x})$ : 
    \begin{align}
        f_i^{\lambda}(\bm{x},t) = \sum_{s=0}^{p-1} W(\bm{x}_{\mathrm{dp}}-\bm{x}_s)  f_i^{\lambda}(\bm{x}_s,t-\Delta t)
        \label{eq:propagationInterpolated}
    \end{align}    
{In this work we use a 2nd-order Lagrangian interpolation stencil together with a k3 limiter for this purpose \cite{ReyhanianPHD}.} 

In practice, populations at each lattice node are stored at their own unique local reference frame, $\lambda_s$, and prior to interpolation the populations in the interpolation kernel need to be transformed from their local reference frames to a common reference frame, $\lambda$, the reference frame of the node. 
In order to ensure consistent advection and impose Galilean invariance, frame transformations are designed to be moment invariant i.e a change in reference frame of a population does not cause any change in the macroscopic moments. 

We define such a moment-invariant transformation from frame $\lambda_s$ to frame $\lambda$ : 
\begin{equation}
    f_i^{\lambda} = \mathcal{G}_{i\lambda_s}^{\lambda} (f^{\lambda_s})
    \label{eq:transformationSymbolic}    
\end{equation}

Combining equations \ref{eq:propagationInterpolated} and \ref{eq:transformationSymbolic}, we get the complete equation for semi-Lagrangian advection with frame transformation: 
    \begin{align}
        f_i^{\lambda}(\bm{x},t) = \sum_{s=0}^{p-1} W(\bm{x}_{\mathrm{dp}}-\bm{x}_s)  \mathcal{G}_{i\lambda_s}^{\lambda} (f^{\lambda_s}(\bm{x}_s,t-\Delta t))
        \label{eq:fullAdvect}
    \end{align}    

An unknown that yet remains is the reference frame $\lambda$. Since the frame depends on the local velocity and temperature, it is unknown a-priori to the advection process. In previous works, an iterative predictor-corrector methodology was employed, where starting with a guess, the frame was updated until a given convergence was achieved. 

Here, we instead arrive at the local reference frame using interpolation from the neighbouring nodes. As reported by Hosseini et. al.\cite{Hosseini_2020}, the errors are very small for shift velocities close to the local velocity. 
While interpolation does not ensure an exactly co-moving reference frame, we expect the interpolated frame to be close to the exact reference, and indeed this is what is observed in simulations on comparison between an exactly co-moving frame obtained iteratively and a reference frame obtained via interpolations.    

\subsubsection{Frame Transformation}
In the previous section we have introduced the frame transformation,  $\mathcal{G}_{i\lambda_1}^{\lambda_2}$, as a black box by just outlining its output. 
Here we go into more detail about the frame transformations that are a significant step in the PonD algorithm. We follow the work of Sawant et al\cite{sawantPonD_2022} and Zipunova et al\cite{Zipunova_Reg2021}.

We define the set of moments for given frame $\lambda = (\boldsymbol{u},T)$
\begin{align}
    \sum_{i=0}^{Q-1} f_i v_{ix}^p(u_x,T) v_{iy}^q(u_y,T) v_{iz}^r(u_z,T) &= M^{\lambda}_{x^p y^q z^r},
    \label{eq:genericMomentUptoO3}
\end{align}
up to order $N  = p + q + r$. The distribution functions $f_i$ are constructed using the Grad's Hermite polynomial expansion  \cite{grad_kinetic_1949}.
\begin{align}
    &f_i = w_i \sum_{n=0}^{\infty} \frac{1}{n!} a(\bm{m};\lambda(\bm{u},T))^{(n)} H_i^{(n)}.
    \label{eq:gradGeneral}
\end{align}

Here, $a(\bm{m};\lambda(\bm{u},T))^{(n)}$ are the coefficients constrained by the above moments and $\bm{m}$ is the vector of $n_m$ moments of the distribution functions. In this work we ensure that moments of order up to $N=3$ are satisfied. 

As mentioned previously, the main property of the frame transformations is the invariance of macroscopic moments, irrespective of frame of reference. Writing the vector of $n_m$ moments of the distribution functions $f_i^\lambda$ as $\bm{m}(\bm{f}^{\lambda})$, we have:
\begin{align}
    \bm{m}(\bm{f}^{\lambda})=\bm{m}({\bm{f}^{\lambda'}}).
     \label{eq:momentInvariance}
\end{align}

For constructing populations in the frame $ \lambda^{'} $, we exploit equation \ref{eq:momentInvariance} and the known moments in frame $ \lambda $ together with equation \ref{eq:gradGeneral}: 
\begin{align}
    &f_i^{\lambda'} = w_i \sum_{n=0}^{\infty} \frac{1}{n!} a(\bm{m}^\lambda;\lambda'(\bm{u},T))^{(n)} H_i^{(n)}
    \label{eq:gradGeneralTransform}
\end{align}

This is the exact operation that is short-handed as $\mathcal{G}_{i\lambda}^{\lambda'}$ 

\section{ \label{sec:BoundaryCondtions} Treatment of Boundary Condition}

Boundary treatment in lattice Boltzmann derived methods typically depends on bounce-back mechanism near wall to identify and populate missing populations close to the fluid-solid interface. From simple bounce-back and its derived schemes \cite{Bouzidi2001,LALLEMAND2003406,Yu2012,Ginzburg2003}, to higher order schemes such as Grad's \cite{dorschner2015_grad}, the uniformity of the stencil throughout the domain is used to systematically apply the same steps to determine the missing populations. 
The typical algorithm is : 
\begin{enumerate}
    \item Identify missing populations, whose departure points lie on lattice nodes that are outside the fluid bounds.  
    \item Using bounce back rules or a more accurate gradient preserving interpolation, calculate the populations which are missing. \item In case of interpolation, linear/bi-linear interpolation is straightforward as all the points in the stencil are on lattice due to lattice stencil isotropy. %
\end{enumerate}

In PonD however, due to the scaling of the lattice velocities at each node, the isotropy of the lattice is lost and together with the loss of on-lattice advection due to shift, bounce-back and simple interpolation based boundary treatments become inadmissible. 

\begin{figure}[h!]
    \centering
    \includegraphics[width=\linewidth]{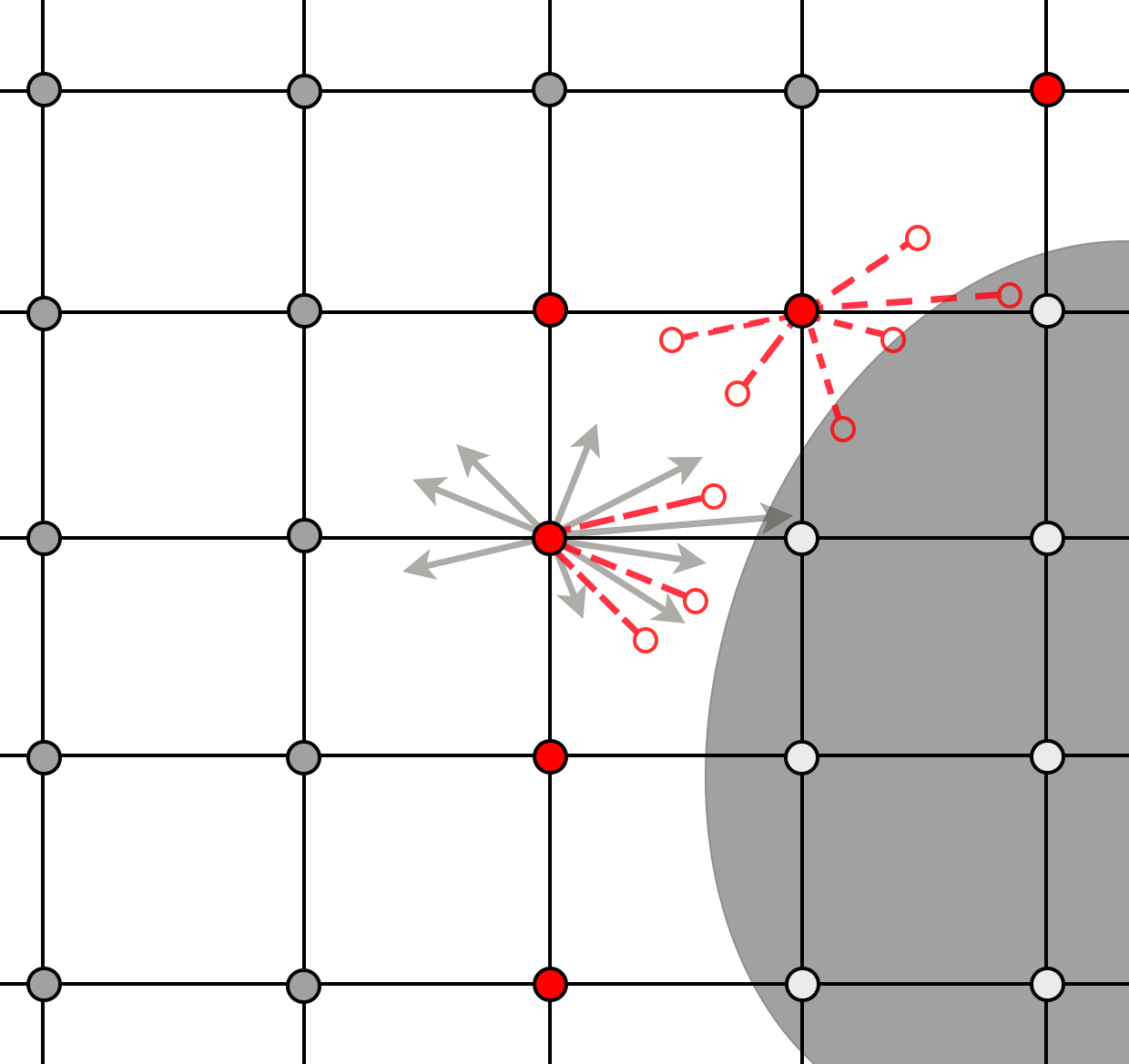}
    \caption{Boundary nodes and missing interpolation stencil at boundary. A-priori information about missing populations not available. Missing populations due to  incomplete stencil even if departure point (open circle) lies within fluid domain. }
    \label{fig:pond_boundary}
\end{figure}

The schematic in figure~\ref{fig:pond_boundary} represents how such a scenario looks like in practice. By definition, the stencil (and therefore, the missing populations) depend on the local velocity and temperature. It is therefore possible for populations with departure points that lie within the fluid to be missing, owing to an incomplete interpolation stencil during advection. Moreover, these departure points and whether the populations are missing cannot be determined a-priori. The departure point now also depends on the local velocity and temperature whereas in standard off-lattice LB methods it only depends on the fixed discrete velocities, $c_i$

Therefore, in this work, we mark all lattice nodes with an incomplete interpolation stencil to be boundary nodes, replacing all populations. At these boundary nodes, we follow a non-equilibrium extrapolation \cite{Guo2002,FengSagaut_2019} approach by splitting the boundary populations into an equilibrium and non-equilibrium part. We use Dirichlett and Neumann values of macro quantities to impose the equilibrium distribution and interpolate from available fluid neighbours the non-equilibrium part. 
\begin{align}
    f_i =  f_i^{eq}(\rho_{bnd}, u_{bnd}, T_{bnd}) + f_i^{(1)}
    \label{eq:BasicGuoEq}
\end{align}

To obtain a robust interpolation stencil for the boundary nodes irrespective of the boundary, we use inverse distance interpolation, which does not require a symmetrical stencil.

\begin{figure}[t]
    \centering
    \includegraphics[width=\linewidth]{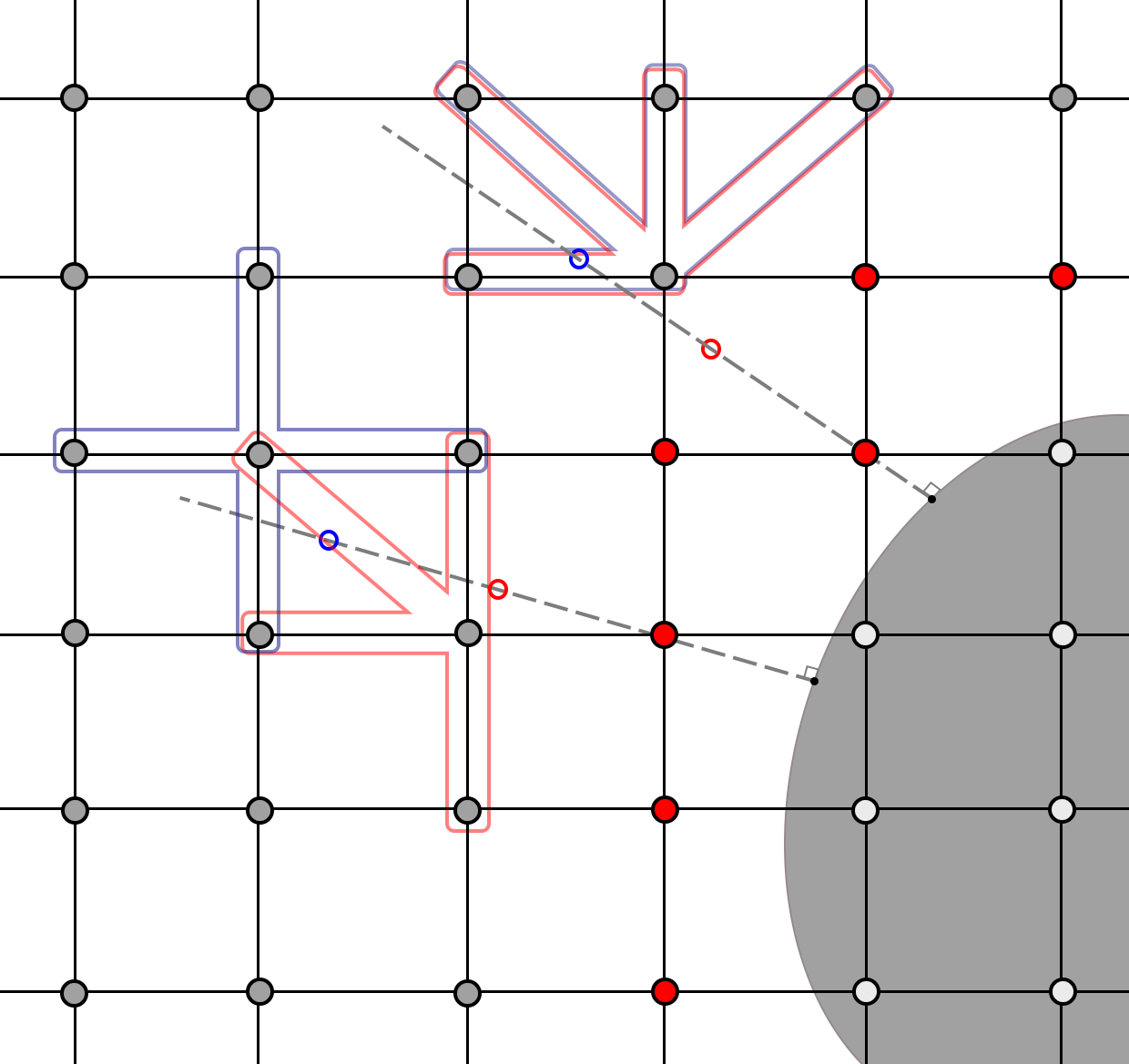}
    \caption{Interpolation at boundary nodes (filled circle) along normal and interpolation stencil for computing reference values (open circles) using inverse distance weighting. }
    \label{fig:pond_IDW}
\end{figure}

The interpolation consists of two steps and can be seen in the schematic in figure~\ref{fig:pond_IDW}. 
The value at the boundary node (filled red circle) is linearly interpolated from the known value at the wall, and at two reference points (open red and blue) in the fluid that lie along the normal from the wall to the boundary node.   

For Dirichlet boundary conditions we have,
\begin{align}
    \phi_{bnd} = \frac{2dw}{dw+\delta_x}\phi_1 + \frac{-dw}{dw+2\delta_x}\phi_2 + \frac{2\delta_x^2}{(dw+\delta_x)(dw+2\delta_x)} \phi_{wall}
    \label{eq:bndInterp}
\end{align}

Similarly for Neumann boundary conditions i.e $ \frac{\partial \phi_{wall}}{\partial n} = 0 $, we  have :
\begin{align}
    &\phi_{bnd} = \frac{4(dw+\delta_x)\phi_1 -(2dw+\delta_x)\phi_2 }{2dw+3\delta_x }
    \label{eq:bndInterpNeumann}
\end{align}

The value at each off-lattice reference point (open red and blue circles) is computed using inverse-distance interpolation. 
The stencil for each is obtained by searching for the nearest admissible (i.e fluid/non-boundary) neighbour node in a clock-wise fashion. In this work, a stencil consisting of 8 nodes was used. 

For an off-lattice reference point $\bm{x_p}$, with $N_n$ nearest neighbours, the interpolation using inverse distance weighting can be done as: 
\begin{align}
    &\phi_p = \sum_{i=1}^{N_n} \frac{ \Vert \bm{PI} \Vert^{-k} }{ \sum_{i=1}^{N_n}  \Vert \bm{PI} \Vert^{-k}  },
    \label{eq:IDW}
\end{align}
where $\Vert \bm{PI} \Vert $ is the Euclidean distance between the reference point, $\bm{x_p}$ and the neighbour, $\bm{x}_i$. 

Combining the above, the following algorithm is followed at each boundary node:  
\begin{enumerate}
    \item The values of the macroscopic variables, $\rho, u, T$, are computed at the two reference points using inverse-distance weighting from equation \ref{eq:IDW}.
    \item The macroscopic values at the boundary node are interpolated along the normal from equation \ref{eq:bndInterp} or \ref{eq:bndInterpNeumann}.
    \item The equilibrium part of the boundary distribution (equation \ref{eq:BasicGuoEq}) is computed using equation \ref{eq:Exactequilibrium}.
    \item For the non-equilibrium part, the same process is repeated. First the non-equilibrium population at reference points is calculated using inverse distance interpolation. 
    \item The non-equilibrium population at the boundary node is then interpolated from the reference points using equation \ref{eq:bndInterpNeumann}.   
\end{enumerate}
  { We note here that care must be taken while interpolating populations, to first transform the populations in the stencil to the reference frame of the boundary nodes, $\lambda_{bnd} = \{u_{bnd},T_{bnd}\}$ }.

\section{\label{sec:results}Results}
In the following section we demonstrate the suitability of aforementioned boundary conditions to simulate high speed compressible flows using several two-dimensional test cases that exist in literature. We start with the canonical case of flow over cylinder to establish a sanity check. We then move on to compressible test cases with flow over a wedge in the Schardin's problem. Next, we ran multiple simulations over a cylinder in the supersonic regime. In the free-stream, we observe the resulting shock standoff distance. Finally, we show the test case of a cylinder inside a reflecting channel. 
\subsection{Low Speed Flow Over Cylinder}

\begin{figure}[h!]
  \begin{subfigure}[c]{\linewidth}
            \centering\includegraphics[ trim=250 100 0 100, clip, width=\linewidth]{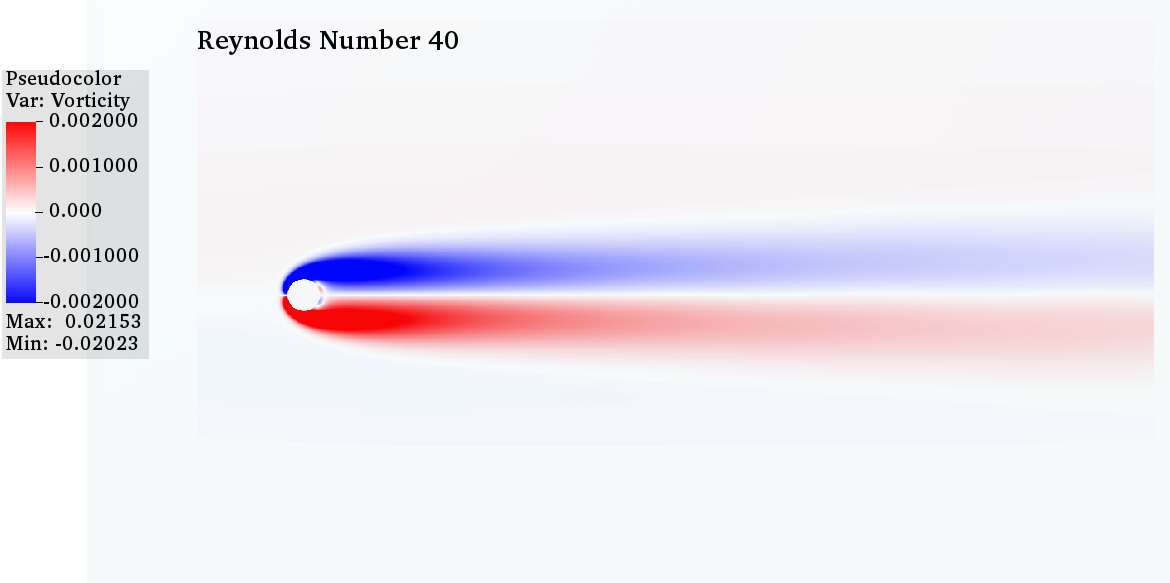}
    \end{subfigure}
    \begin{subfigure}[c]{\linewidth}
            \centering\includegraphics[ trim=250 100 0 100, clip,width=\linewidth]{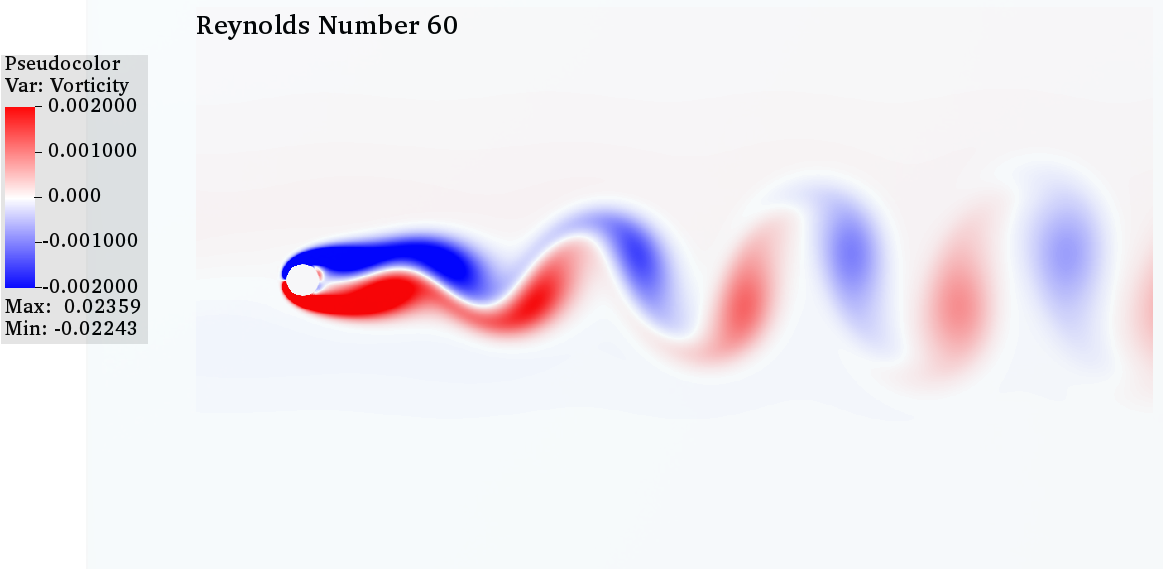} 
    \end{subfigure}
    \begin{subfigure}[c]{\linewidth}
            \centering\includegraphics[ trim=250 100 0 100, clip, width=\linewidth]{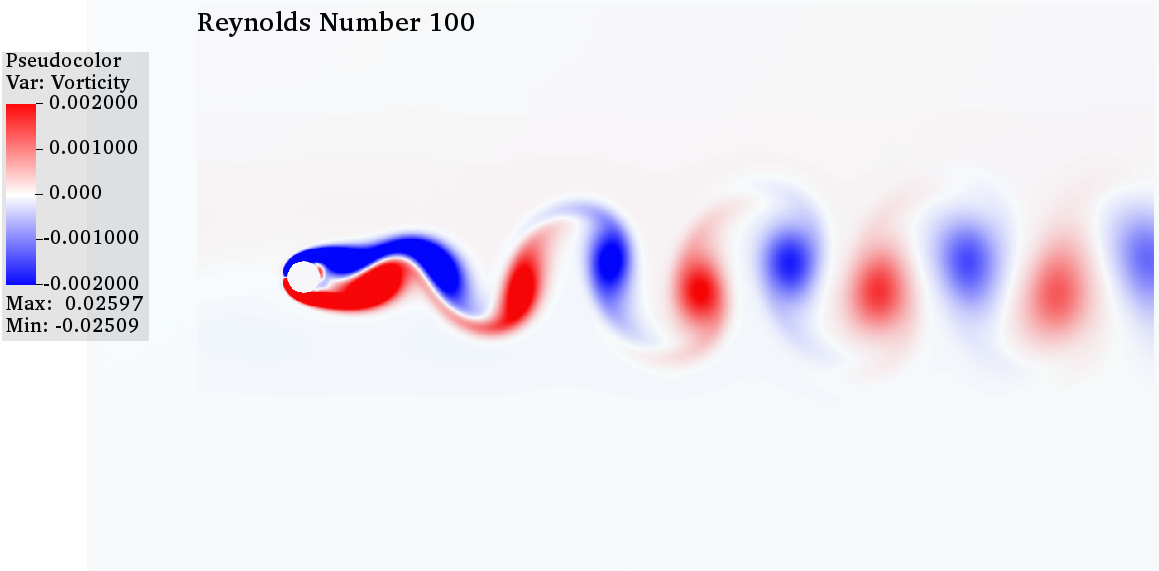} 
    \end{subfigure}
\caption{\label{fig:lowRe_vorticity} Vorticity contours for low speed incompressible flow. Top : $\mathrm{Re= } 40$, Middle : $\mathrm{Re= } 60$, Bottom : $\mathrm{Re= } 100$.  }
\end{figure}

Flow over a cylinder in two dimensions is a widely studied case in fluid dynamics. Its simplicity together with its applications combine to ensure that it has been the focus of many experimental and computational studies over the years. Here, we present low speed flow over a cylinder over various Reynolds numbers and use it as a basic check of our boundary conditions. We resolve the boundary using a resolution of 40 lattice nodes per diameter. 
We observe the critical Reynolds number, the vortex shedding frequency and the length of the re-circulation region in the cylinder's wake. 

Figure ~\ref{fig:lowRe_vorticity} shows the contours of the vorticity and we can observe the onset of vortex shedding as the Reynolds number is increased. 
The critical Reynolds number, at which vortex shedding commences, is observed to be 43, which agrees well with the range observed in both experiment and simulation \cite{RASTAN2022110393}. 

The frequency of vortex shedding is obtained via FFT of the velocity data in the wake of the cylinder, and reported in table \ref{tab:strouhal} against results from literature. Table \ref{tab:recirculation} reports the re-circulation region lengths for low Reynolds number cases without vortex shedding. In both cases, the results agree well with literature and provide us with a strong basis to move to higher Reynolds numbers and Mach numbers.  

\begin{table}[h!]
\begin{ruledtabular}
\begin{tabular}{ccc} 
\multicolumn{1}{p{2.5cm}}{\centering Reynolds Number \\ Re }  & \multicolumn{1}{p{3cm}}{\centering Strouhal Number \\ PonD }  &   \multicolumn{1}{p{3cm}}{\centering Literature \\ \cite{lienhard1966,NORBERG200357}}\\
\hline \noalign{\vskip 2mm}
60 & 0.15 & 0.14-0.15\\ 
100 & 0.1625 & 0.16-0.17\\
\end{tabular}
\end{ruledtabular}
\caption{\label{tab:strouhal} Strouhal Number  }
\end{table}

\vspace{-0.5cm}

\begin{table}[h!]
\begin{ruledtabular}
\begin{tabular}{ccc} 
\multicolumn{1}{p{2.5cm}}{\centering Reynolds Number \\ Re }  & \multicolumn{1}{p{3cm}}{\centering Re-circulation Length \\ PonD }  &   \multicolumn{1}{p{3cm}}{\centering Literature \\ \cite{Taneda_recirc,Takami_RecircLength,coutanceau_bouard_1977} }\\
\hline \noalign{\vskip 2mm}
20 & 1.78 & 1.7-1.89 \\ 
40 & 4.7 & 4.25-5.7  \\
\end{tabular}
\end{ruledtabular}
\caption{\label{tab:recirculation} Re-circulation Length}
\end{table}

\subsection{Schardin's Problem}

The Schardin's problem is a famous test case, first carried out in 1957 \cite{Schardin1957}, showing the interaction of a planar shock front with a wedge body.
The planar shock impinges on the the wedge and the resulting interaction leads to shock reflection and diffraction which results in a complex flow field with tip vortices, triple points, mach stems etc. 

\begin{figure}[h!] %
   \begin{subfigure}[c]{\linewidth}
            \includegraphics[trim=0 250 0 60, clip,width=0.96\linewidth]{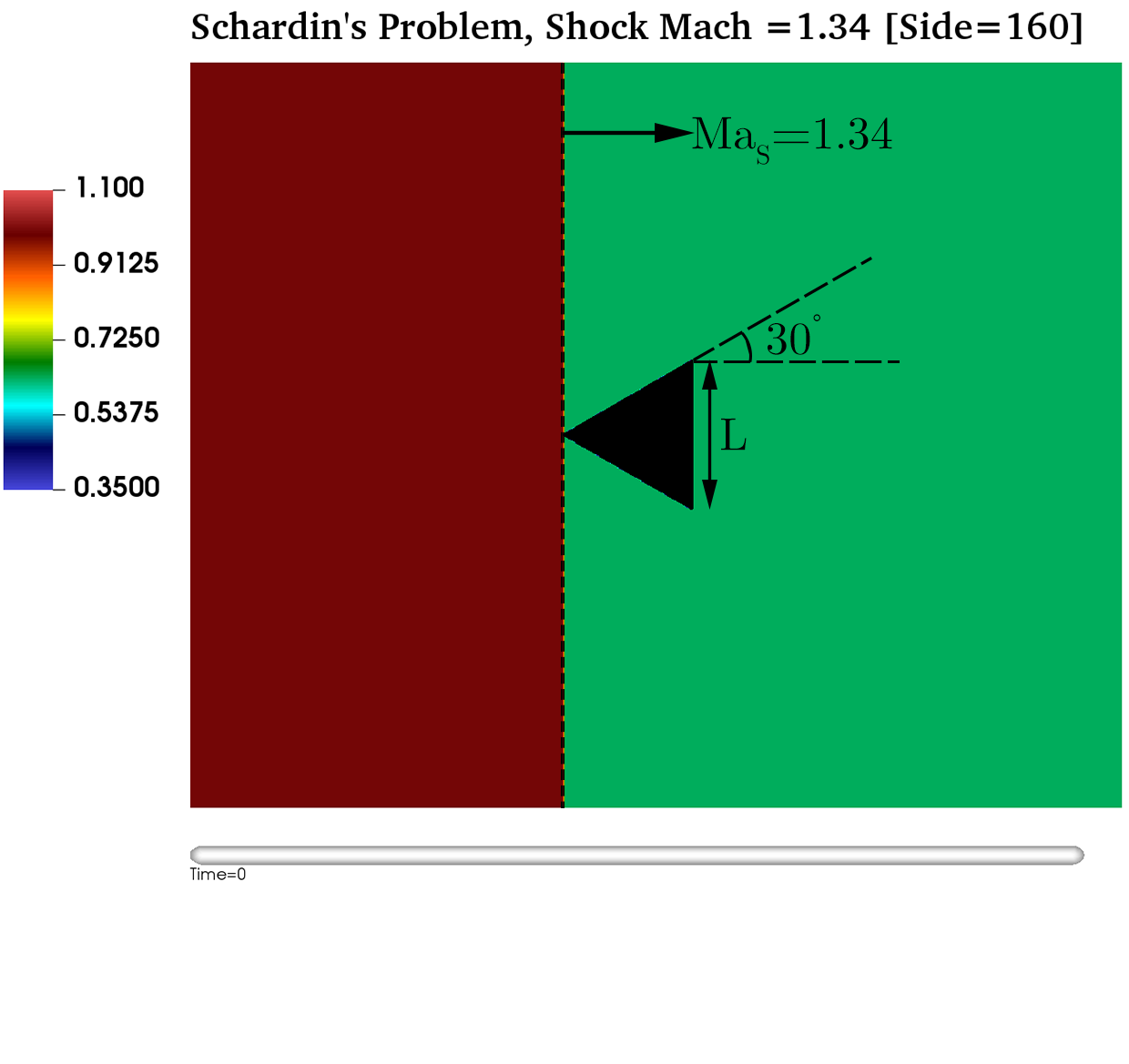} \hspace*{\fill} %
    \end{subfigure}

   \begin{subfigure}[c]{0.49\linewidth}
            \centering\includegraphics[trim=260 130 20 55, clip,width=\linewidth]{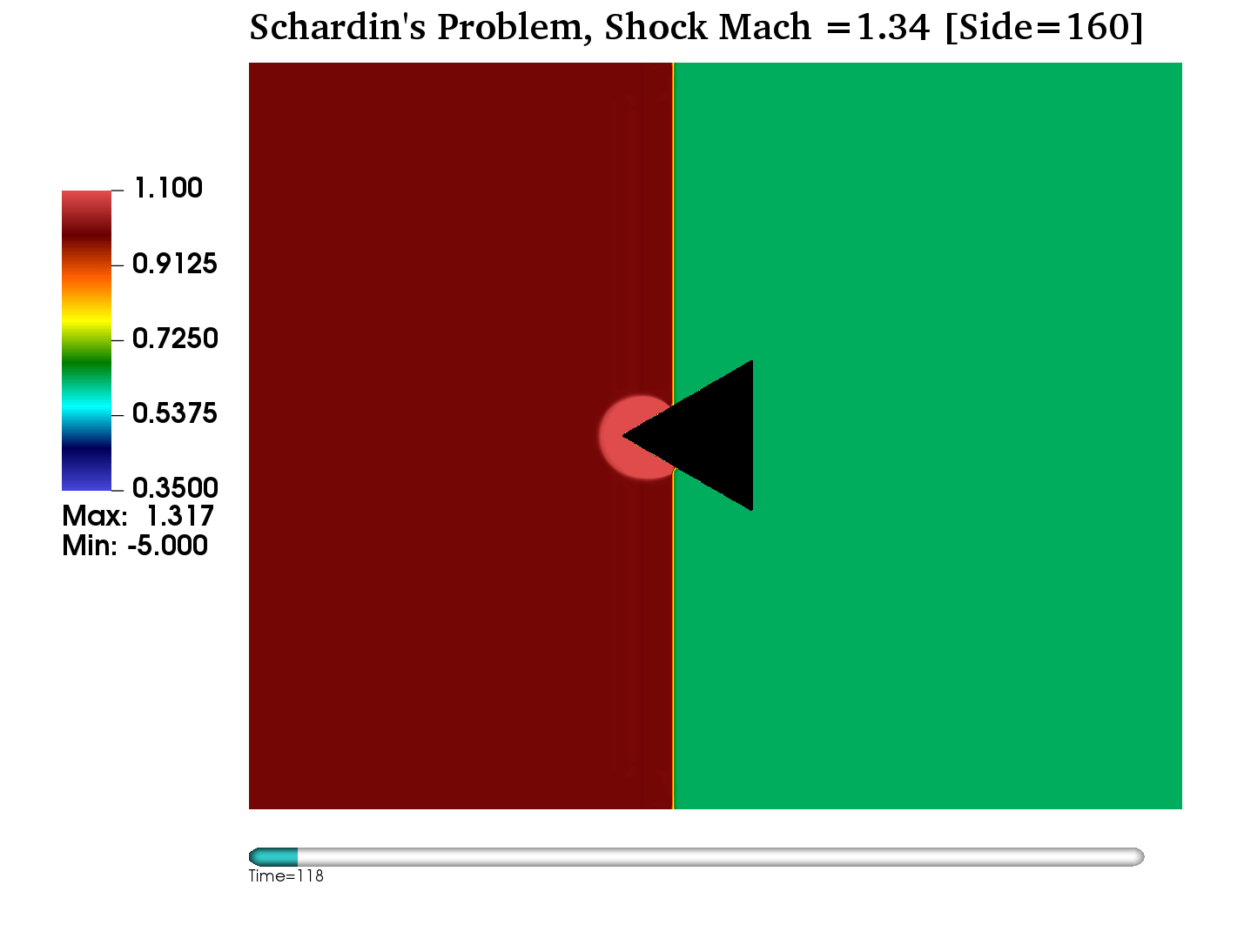} %
    \end{subfigure}
    \begin{subfigure}[c]{0.49\linewidth}
            \centering\includegraphics[trim=260 130 20 55, clip,width=\linewidth]{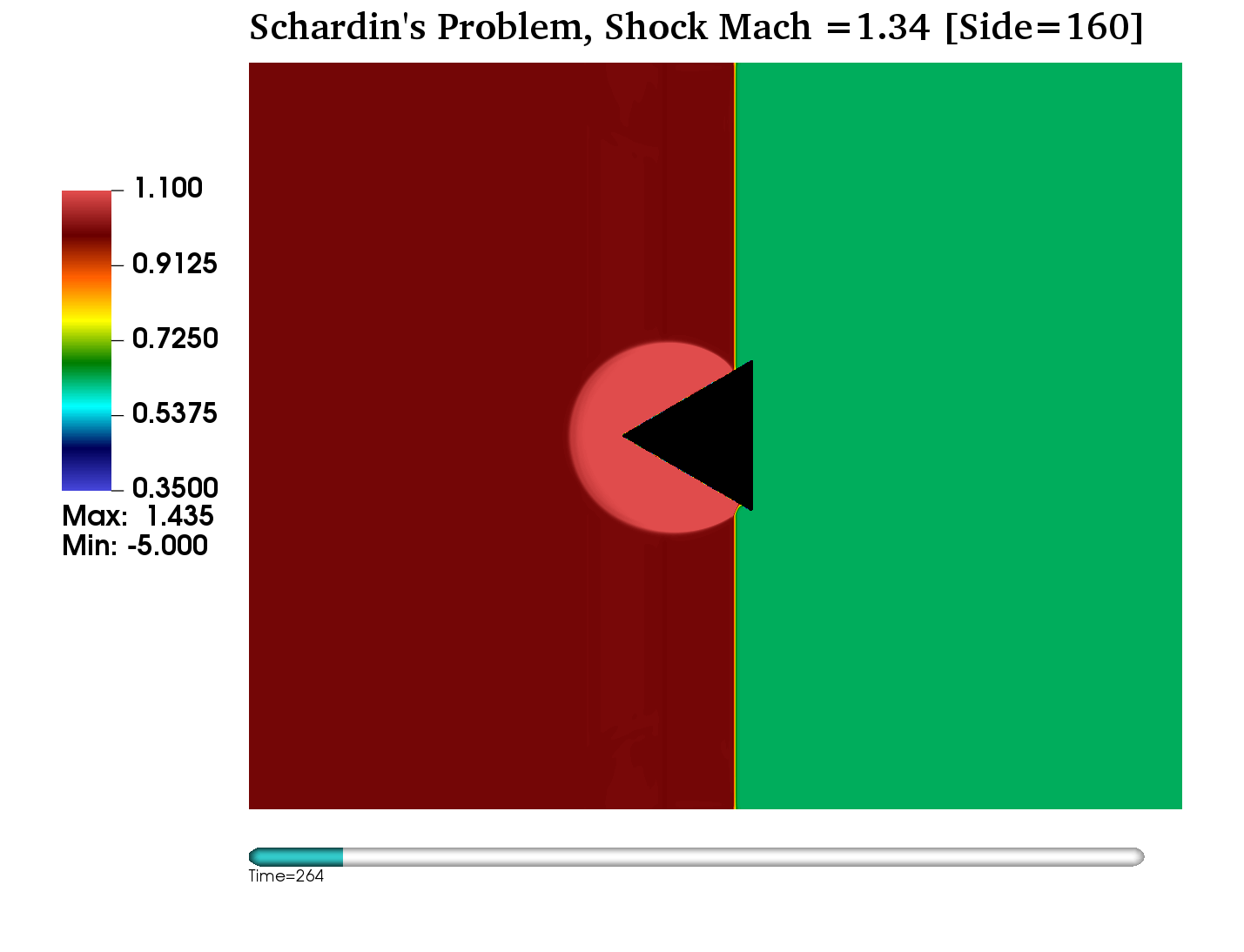} %
    \end{subfigure}
       \begin{subfigure}[c]{0.49\linewidth}
            \centering\includegraphics[trim=260 130 20 55, clip,width=\linewidth]{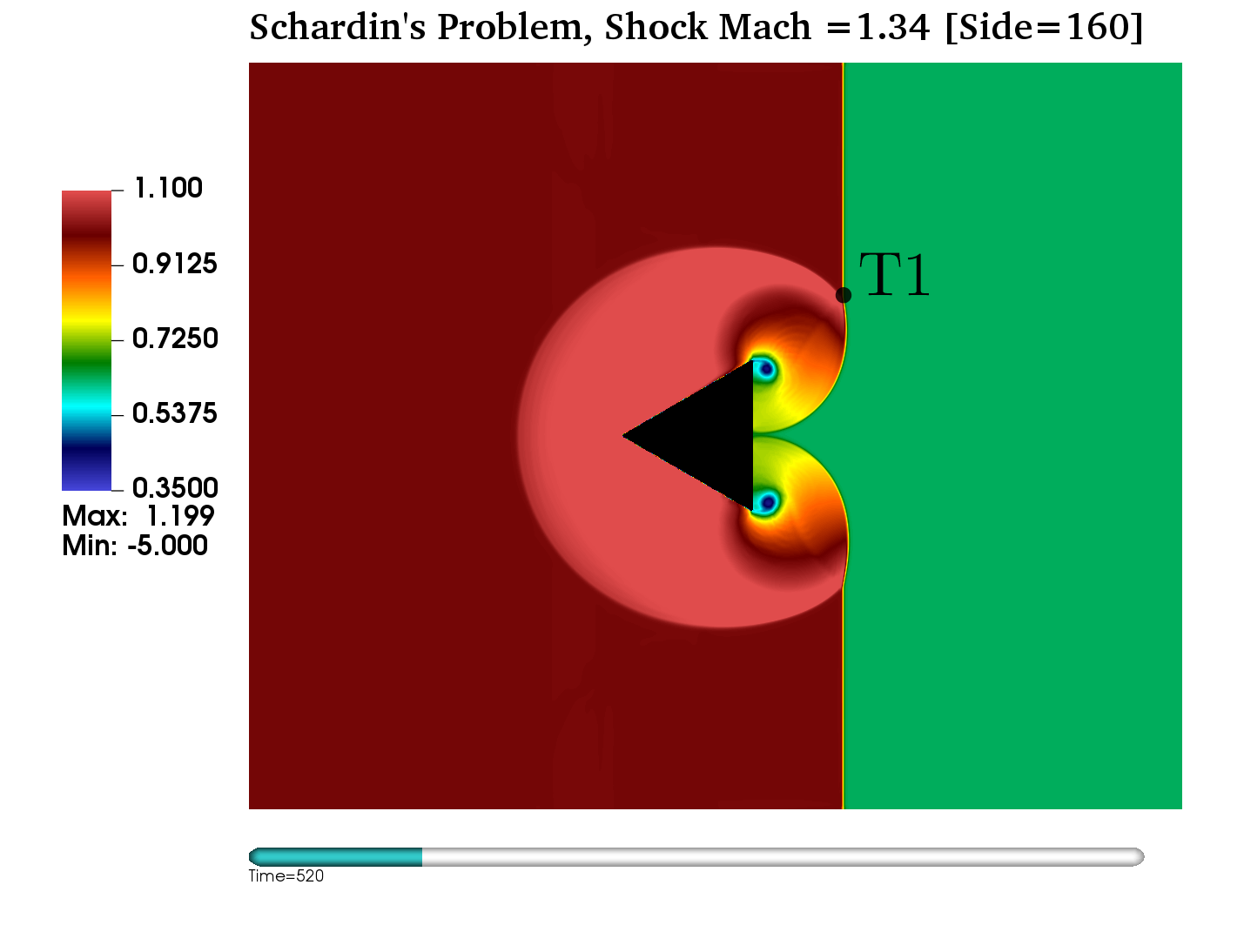} %
    \end{subfigure}
    \begin{subfigure}[c]{0.49\linewidth}
            \centering\includegraphics[trim=260 130 20 55, clip,width=\linewidth]{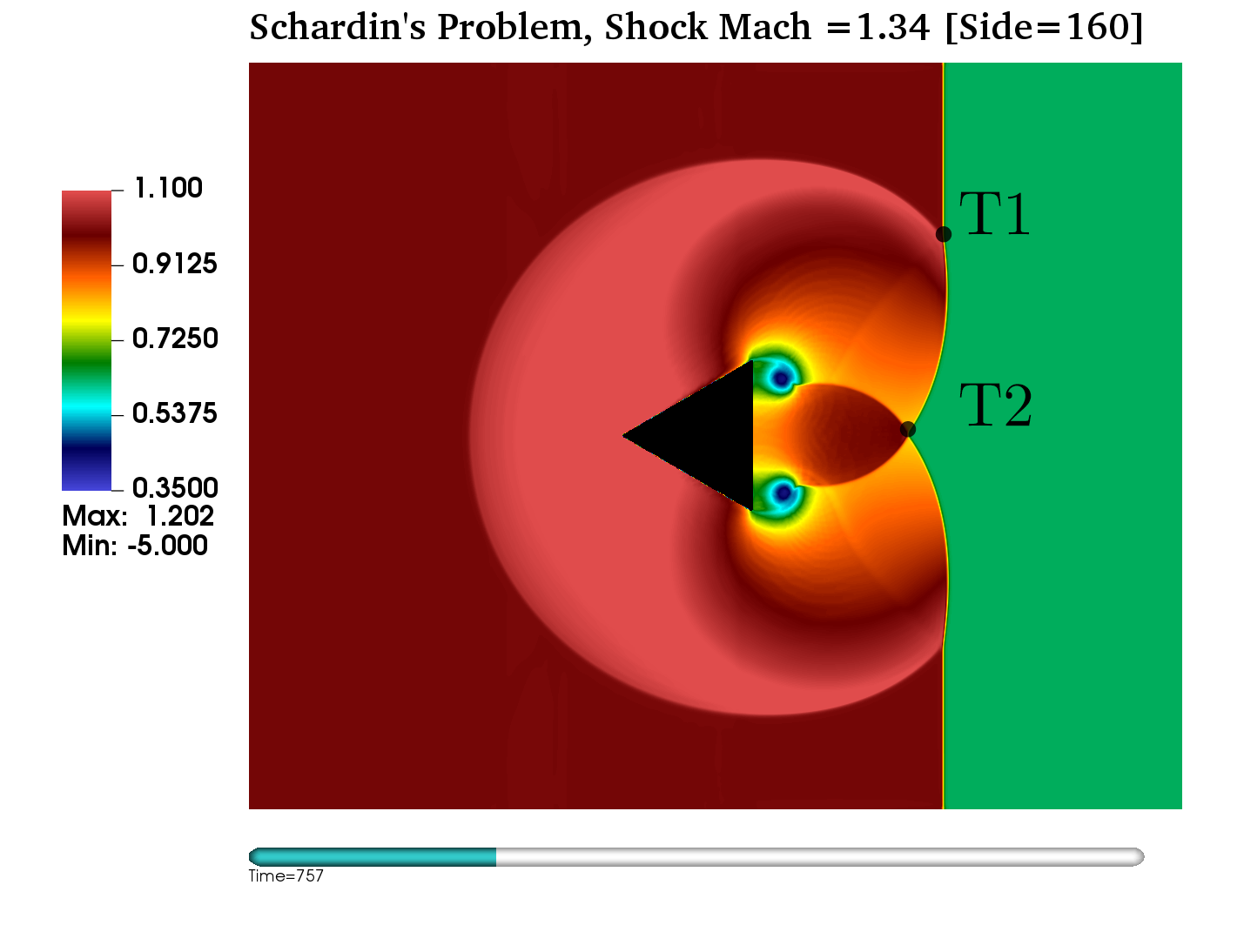} %
    \end{subfigure}

    \begin{subfigure}[c]{\linewidth}
            \centering\includegraphics[trim=260 130 20 55, clip,width=\linewidth]{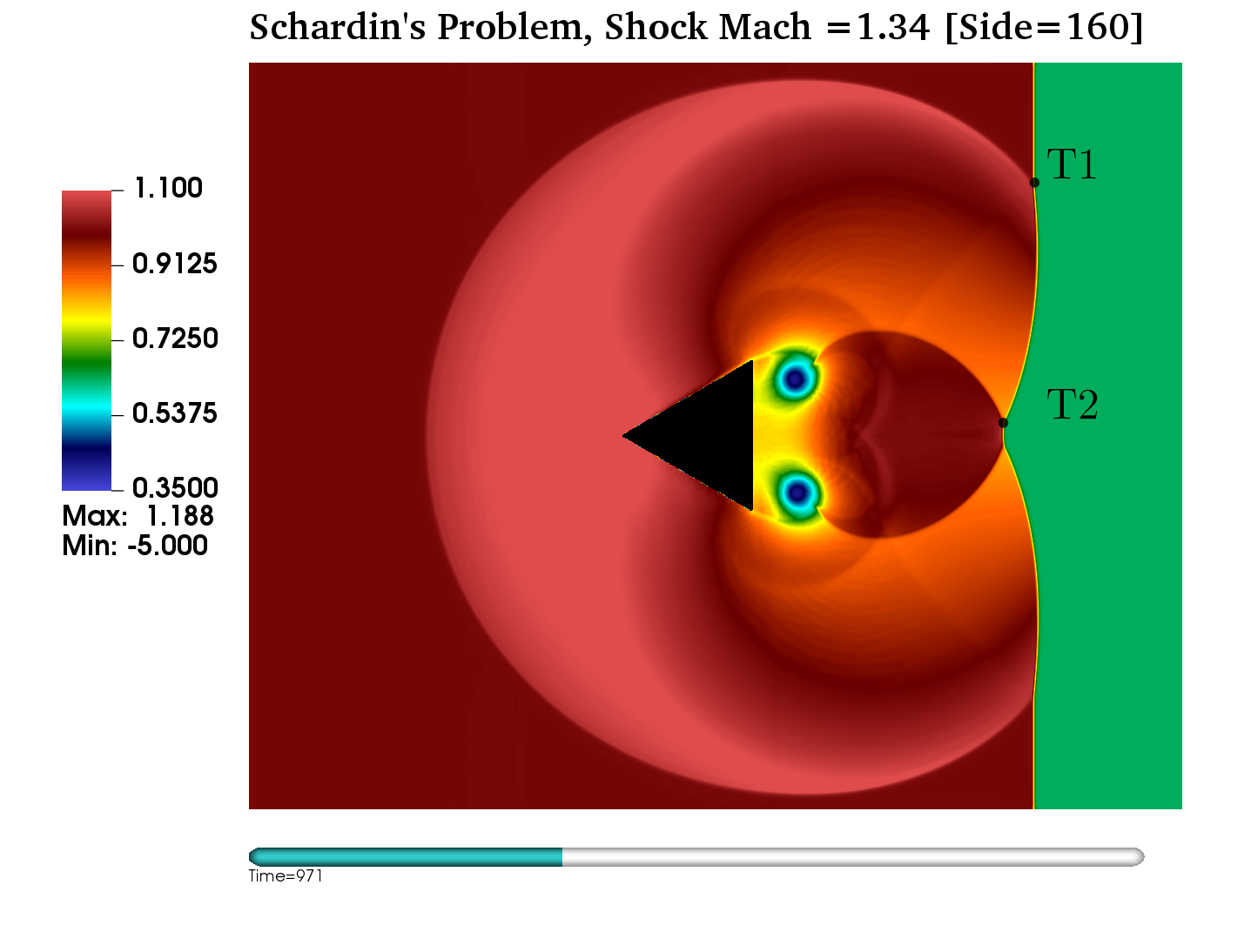} %
    \end{subfigure}
\caption{\label{fig:schardin_rho} Schardin's Problem: Density contours showing progression of planar shock wave over the wedge, shock reflection, formation of diffraction wave, and shock vortex interaction in the wake.}
\end{figure}

Since the first experiment, the test case has been widely studied both experimentally and computationally \cite{Chang2000,Frapolli2016,Kofoglu2022}. We follow the setup of Chang and Chang \cite{Chang2000} who used an equilateral triangle and a shock Mach number, $M_s = 1.34$. The schematic is shown in figure \ref{fig:schardin_rho} and the boundary was resolved using 160 lattice nodes per side. 

\begin{figure}[h!] %
\centering\includegraphics[width=\linewidth]{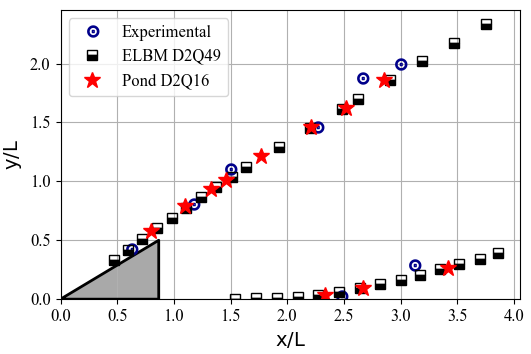}%
\caption{\label{fig:schardin_bench} Schardin problem: Position of triple points.}
\end{figure}

For validation we follow the location of the two triple points arising from the interaction of the planar shock and the reflected shock. The observed location of the triple points has been plotted in figure \ref{fig:schardin_bench} together with results from experiments \cite{Chang2000} and numerical results from an entropic lattice Boltzmann scheme using a D2Q49 lattice \cite{Frapolli2016}. 

Figure~\ref{fig:schardin_rho} shows a series of snapshots of density showing the progression of the planar shock over and around the wedge body. On impinging the body, the the reflected circular shock wave flows upstream in the flow field. As the planar shock moves over the wedge corner, the expansion creates a vortex and turns the planar shock leading to to a diffraction wave. The diffracted wave then impinges on the side of the triangle and interacts with the diffraction wave from the other corner. The diffracted shock after the shock-shock interaction impinge the vortex at the wedge vertex. The position of two triple points being monitored, T1 and T2, are also marked in the figure \ref{fig:schardin_rho}.     

\subsection{Supersonic Flow over 2D Cylinder}

The study of supersonic flows around the blunt bodies has been an area of interest to fluid dynamicists for a long time. The forces acting on a body are extremely sensitive to the behaviour of the shock. Even for a simple geometry such as a sphere or a circle, numerical methods to accurately capture the complex flow dynamics are not simple. 
A significant area of interest is the design of re-entry vehicles, which depend on the shape of the vehicle body to slow them down. 

When a blunt body, such as a sphere, is placed in a supersonic flow field, the oblique shock required to turn the flow is not able to stay attached to the body. The resulting phenomena of a curved shock wave away from the surface of the body is called a detached shock or a bow shock. The formation of a bow shock causes a sharp increase in the drag on the body. This property is exploited by re-entry vehicles to decelerate from hypersonic speeds when they enter the earths atmosphere. 

The position of the detached shock is an important design parameter and it depends on many  factors including the shape of the body and the Mach number. Several researchers have worked on studying the effect of Mach number on a 2D cylinder. The shock standoff distance, $L_R$, is a measured as a factor of the cylinder radius, $R$. It measures the distance of the detached shock from the stagnation point of the body. The most accurate  results have been obtained via experiments \cite{Alperin1950,Kaattari1961,Kim1956}, but good agreement has been observed using numerical methods as well \cite{Sinclair_shockStandoff}.

\begin{figure}[h!] %
\centering\includegraphics[width=\linewidth]{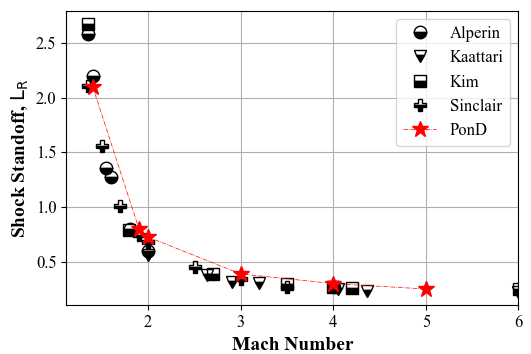}%
\caption{\label{fig:shockStandoff_bench} Supersonic flow over 2D cylinder: Effect of Mach number on shock standoff distance. }
\end{figure}

For this work we ran simulations for a range of Mach numbers between 1 and 5 and observed the shock standoff distance, $L_R$. The boundary resolution was 80 lattice nodes per diameter. The relationship between the shock standoff distance and Mach number is plotted in figure \ref{fig:shockStandoff_bench} together with results from literature. PonD is able to accurately predict the shock standoff distance and the results agree well with literature.  

\begin{figure}[h!] %

   \begin{subfigure}[c]{\linewidth}
            \centering\includegraphics[ trim=0 80 0 80, clip, width=\linewidth]{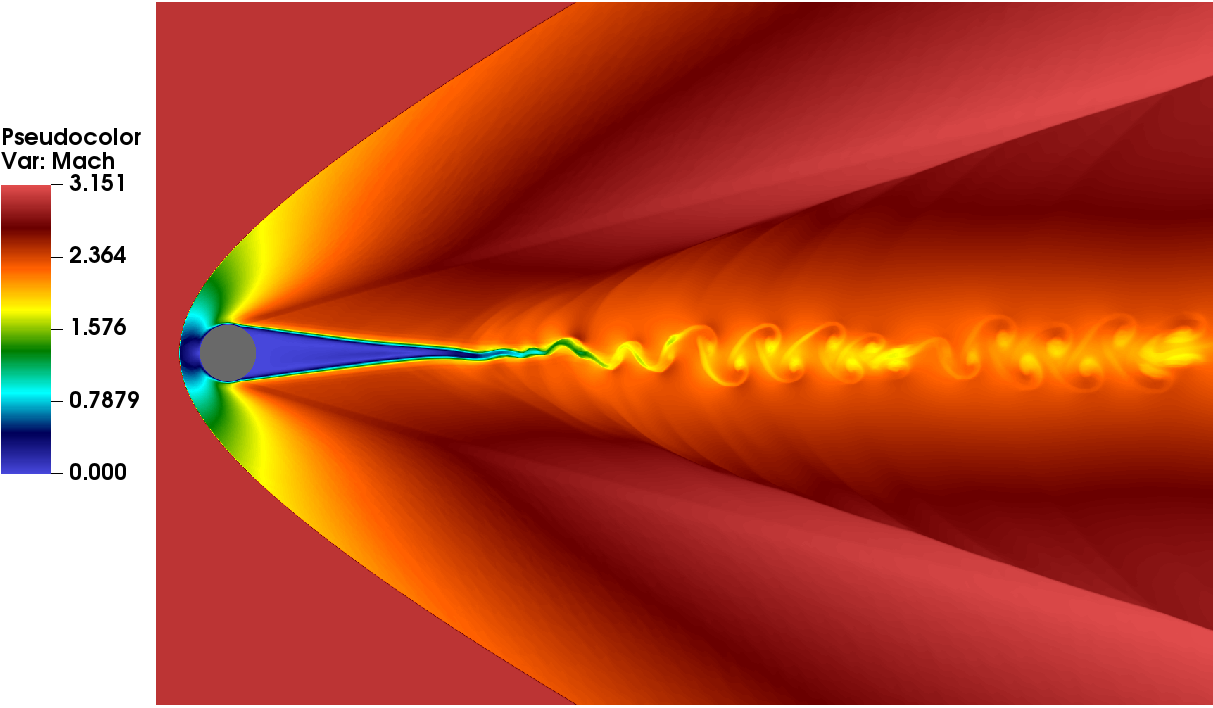} %
    \end{subfigure}
    
    \begin{subfigure}[c]{\linewidth}
            \centering\includegraphics[trim=5 80 0 80, clip, width=\linewidth]{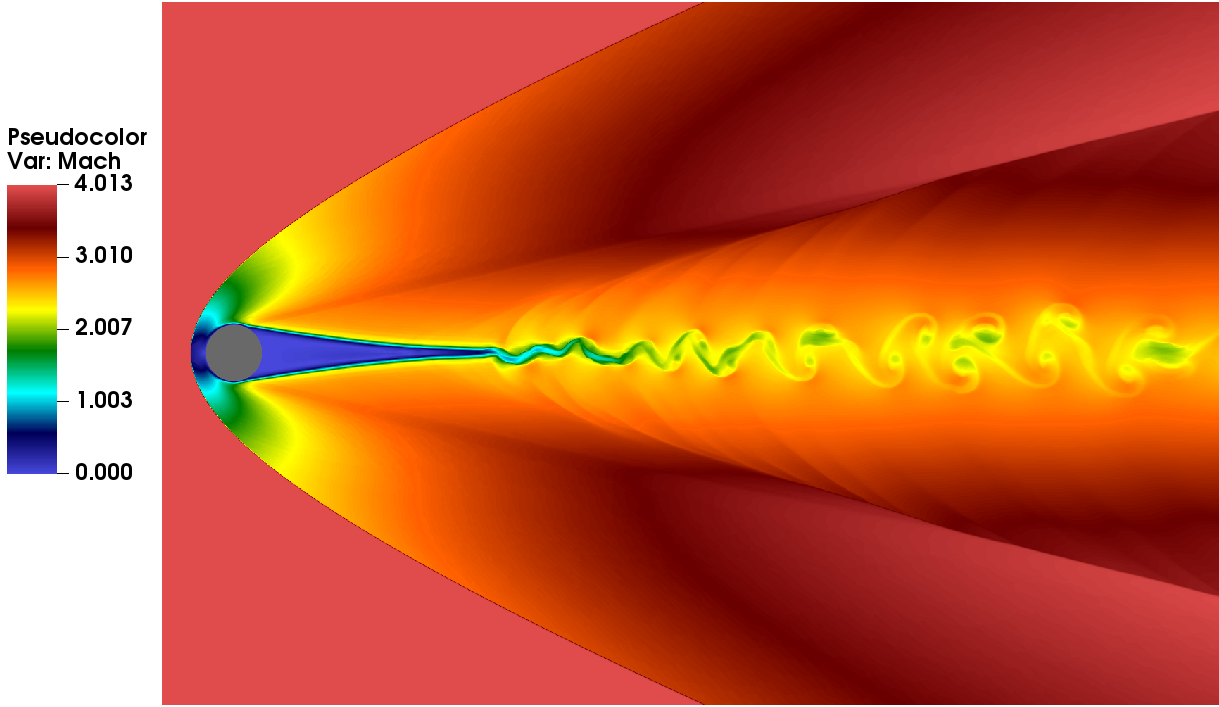} %
    \end{subfigure}
    
   \begin{subfigure}[c]{\linewidth}
            \centering\includegraphics[trim=5 80 0 80, clip, width=\linewidth]{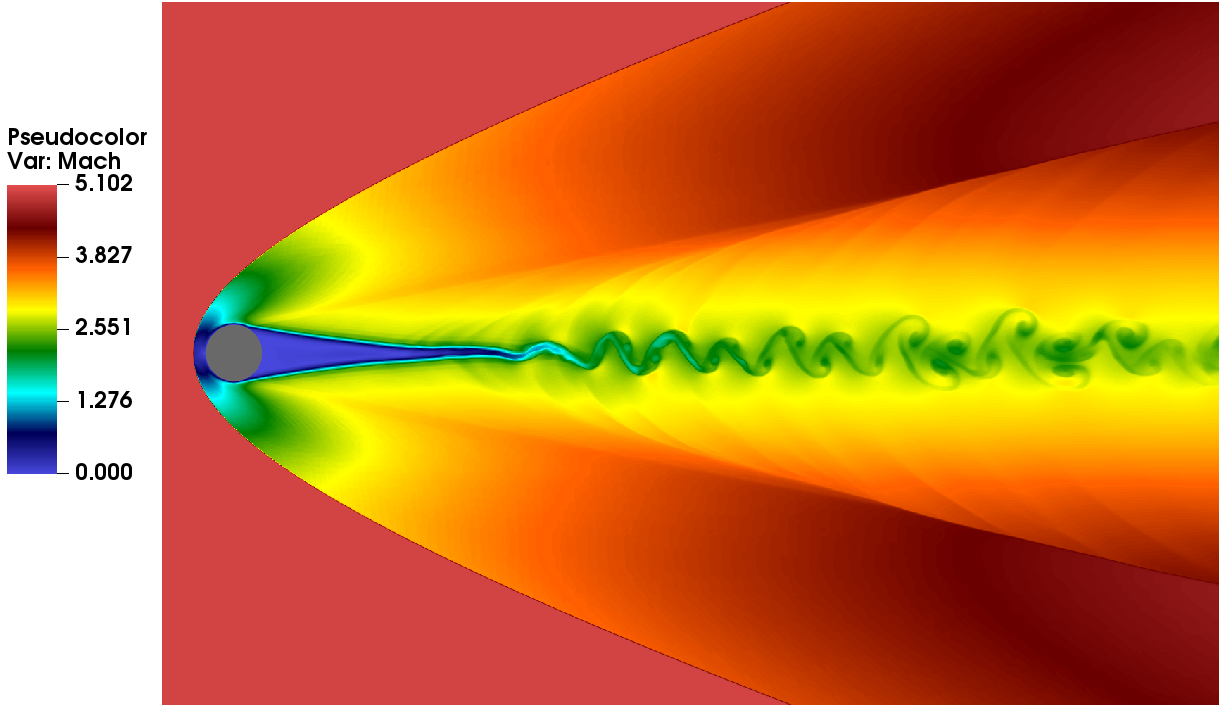} %
    \end{subfigure}
    
\caption{\label{fig:shockStandoff_mach}  Supersonic flow over 2D cylinder: Mach contours showing detached shock upstream of cylinder, from top to bottom Mach 3, Mach 4, and Mach 5. Notice the increasing shock angle with increased Mach number.}
\end{figure}

\begin{figure}[h!] %

   \begin{subfigure}[c]{\linewidth}
            \centering\includegraphics[  width=\linewidth]{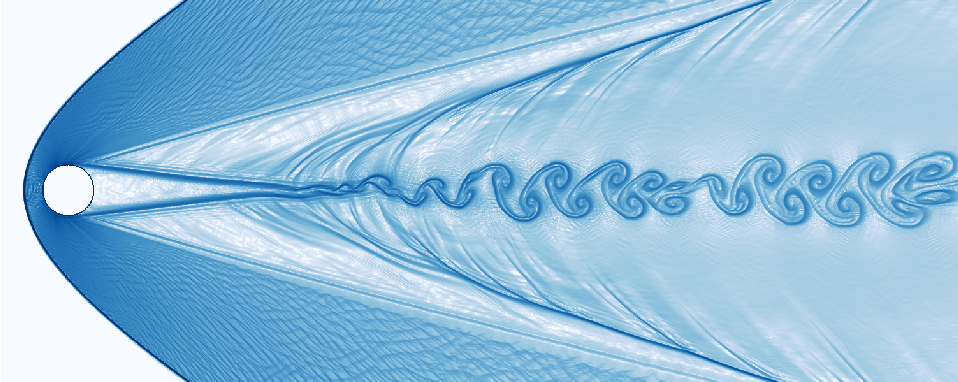} %
    \end{subfigure}
    
    \begin{subfigure}[c]{\linewidth}
            \centering\includegraphics[ width=\linewidth]{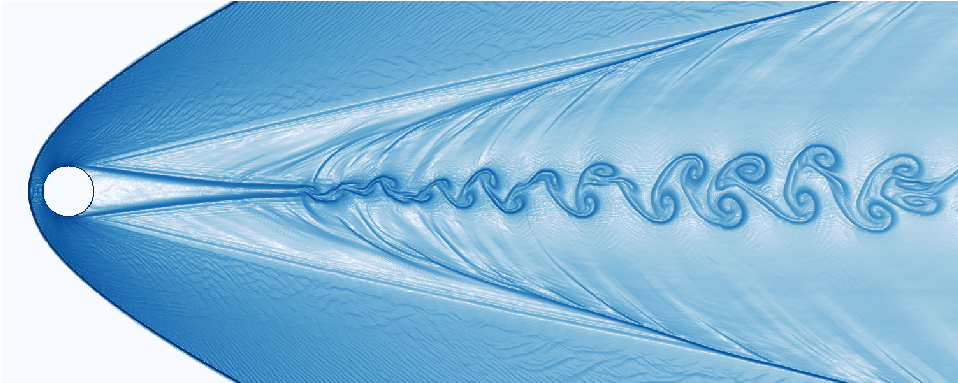} %
    \end{subfigure}
    
   \begin{subfigure}[c]{\linewidth}
            \centering\includegraphics[ width=\linewidth]{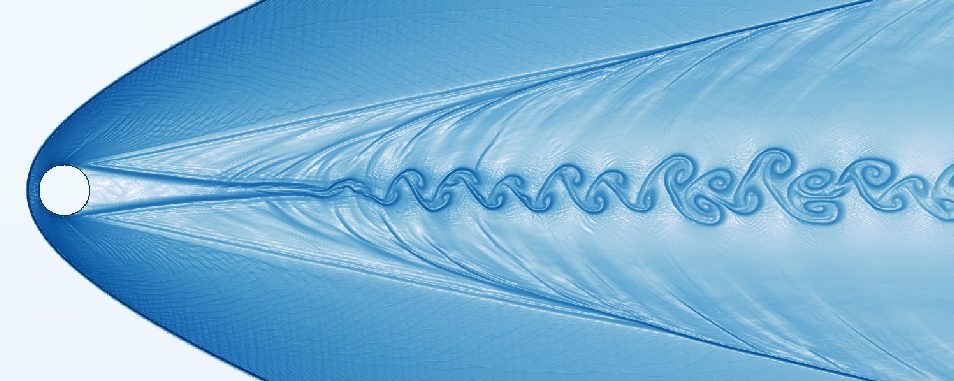} %
    \end{subfigure}
    
\caption{\label{fig:shockStandoff_gradRho}  Supersonic flow over 2D cylinder: Numerical Schlieren showing detached shock upstream of cylinder, from top to bottom  Mach 3, Mach 4, and Mach 5. Notice the vortices arising from the interaction of the secondary shocks with the subsonic region of the wake. }

\end{figure}

The Mach contours in figure \ref{fig:shockStandoff_mach} also offer a qualitative validation for the problem. We observe that with increasing Mach number a corresponding increase in the shock angle (curvature of the bow shock) is seen as well. Moreover, the stagnation zone in front of the body also increases in size with increasing free-stream Mach number. The numerical Schlieren presented in figure \ref{fig:shockStandoff_gradRho} presents another view of the domain. The trailing vortices generated from the interaction of the weak shock downstream of the body with the subsonic wake are clearly visible.

\subsection{Supersonic Flow Over 2D Cylinder Inside a Reflecting Channel}

\begin{figure}[h!]
  \begin{subfigure}[c]{0.95\linewidth}
            \centering
              \includegraphics[width=\linewidth]{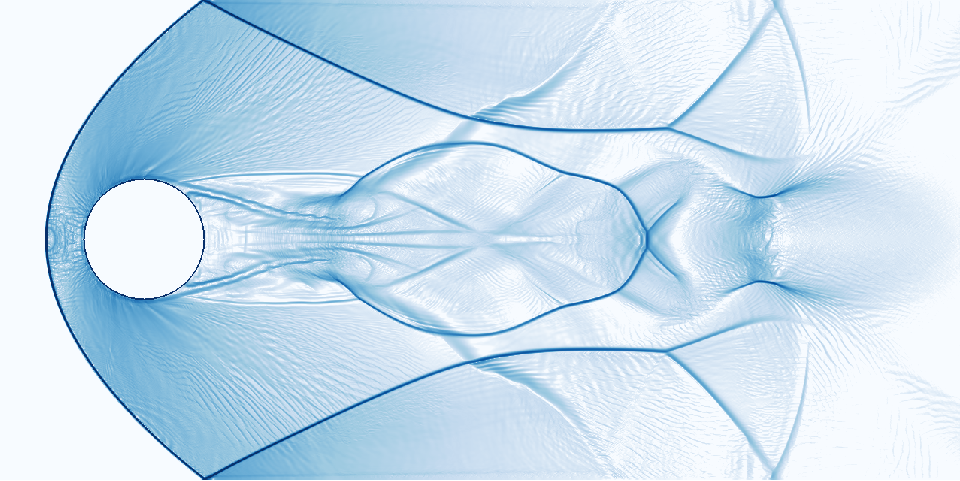}
  \end{subfigure}
  \begin{subfigure}[c]{0.95\linewidth}
            \centering
              \includegraphics[width=\linewidth]{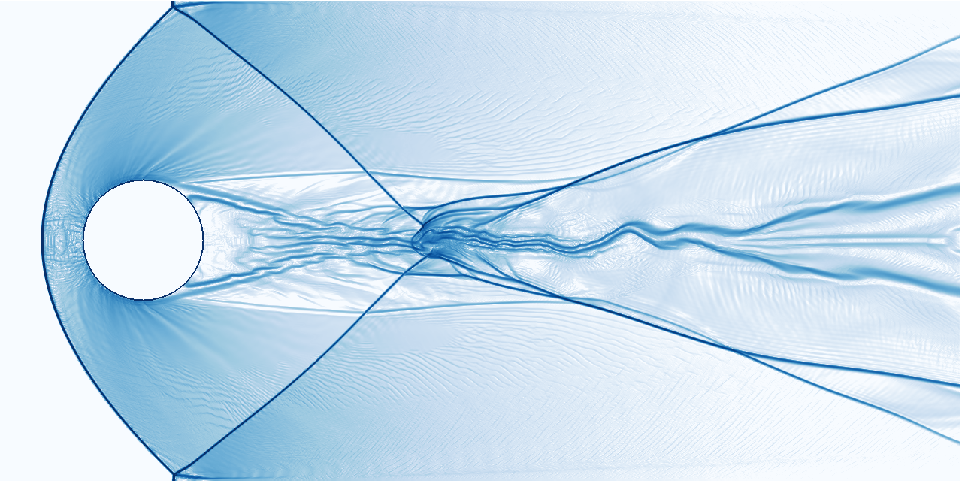}
    \end{subfigure}
      \begin{subfigure}[c]{0.95\linewidth}
            \centering
              \includegraphics[width=\linewidth]{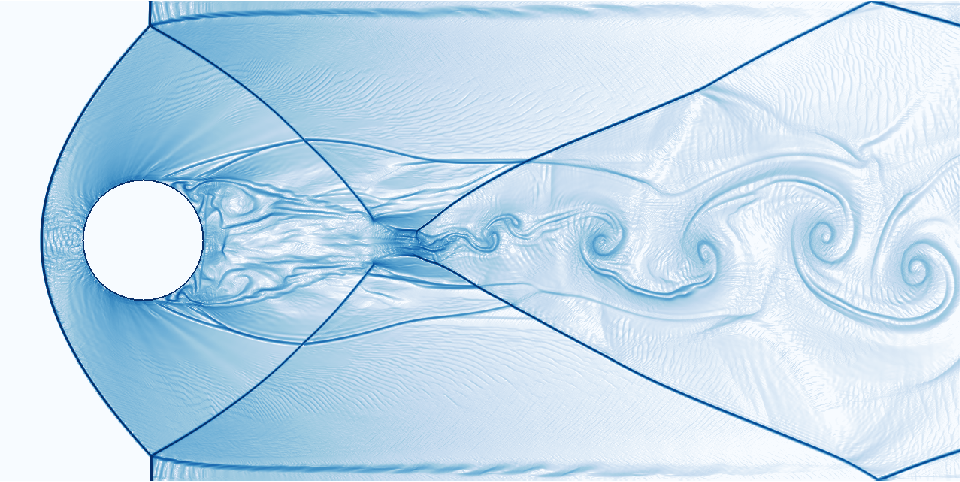}
    \end{subfigure}
    \begin{subfigure}[c]{0.95\linewidth}
            \centering
              \includegraphics[width=\linewidth]{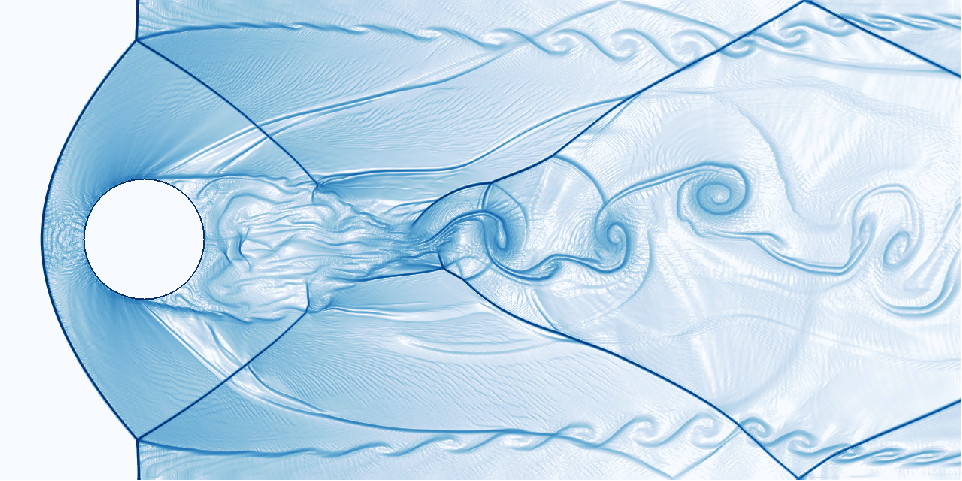}
    \end{subfigure}
    \begin{subfigure}[c]{0.95\linewidth}
            \centering
              \includegraphics[width=\linewidth]{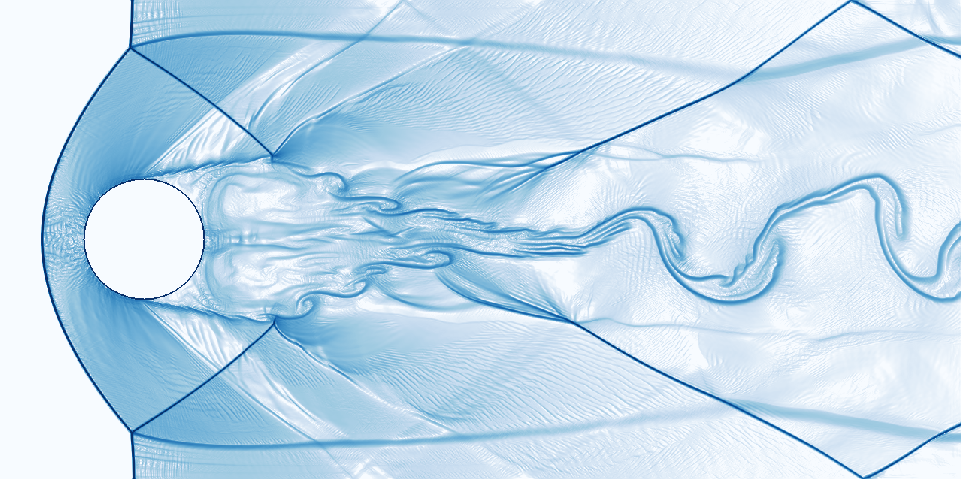}
    \end{subfigure}
    \caption{ Mach 3 flow over a 2D cylinder in a channel: Numerical Schlieren showing formation of bow shock, shock reflection at walls, shock-shock and shock-vortex interaction.}
    \label{fig:rflcwall_gradRho}
\end{figure}

Another interesting test case is that of a circular cylinder inside a closed channel \cite{Nazarov2017,Guermond2018}. 
The channel walls are prescribed with the slip boundary condition ( i.e $ v_{wall} = 0 $ ) while the cylinder has the usual no-slip boundary condition. The supersonic inlet Mach, $ M_{inlet}=3$, and the dimensions of the domain are same as those in Guermond et al \cite{Guermond2018}. The cylinder wall was resolved with 160 lattice nodes per diameter which results in an overall domain size of $1280\times640$. 
The many complex phenomenon that occur in this case make it a good validation of both our model and the boundary conditions. 

As seen in the previous case, the presence of a blunt body in a supersonic flow field leads to a detached shock. Figure \ref{fig:rflcwall_gradRho} presents the numerical Schlieren of the case. Unlike the previous free-stream case, the presence of walls in the vicinity of the body leads to much more complex dynamics. The bow shock impinges on the top and bottom walls and leads to the formation of a triple point whose position settles near the wall after some time. We monitor the position of the triple point and find that it agrees very well with the results from \cite{Guermond2018}.

Further, a Kelvin-Helmholtz instability arises from the triple point and is also captured in our simulations. The strong shocks arising from the wake also reflect from the channel walls and interact with the vortices from the instability. Oblique shocks present on the cylinder cause flow separation and a re-circulation region forms downstream of it. The interaction between the re-circulation region and the reflected shocks from the wall leads to the formation of a complex vortex street, which has been reported in literature, and that is also captured here.   

\section{ \label{sec:conclusions}Conclusions}

We presented a first foray into implementing boundary conditions to simulating complex boundaries at high speeds using particles on demand which is a kinetic theory based approach to simulation of high speed compressible flows.

The challenges involved with computing the missing populations in the semi-Lagranian interpolation stencil were tackled. To achieve this a Guo-like non-Equilibrium extrapolation scheme with inverse-distance weighting was employed for constructing all the populations at the identified boundary nodes.

The boundary conditions were first validated using the canonical case of low Reynolds number flow over a cylinder. The results agreed well with literature. Our scheme was then tested using several simulations of high speed compressible flow and the results were very promising. 

Shock-vortex interaction in the Schardin's problem demonstrated the ability of the implemented boundary condition to adequately resolve sharp corners. The time evolution of the position of the triple point obtained in this work agreed well with the values reported in literature. 

Next, supersonic flow over a 2D cylinder was simulated up to a free-stream Mach number, $M_\infty=5$. Our model was able to accurately predict the distance of the detached shock from the body, as well as capture qualitative details of the flow field including shock vortex interaction in the wake. 

Finally, we tested the case of the 2D cylinder in a channel with reflecting walls. This is a complex case with several shock-shock and shock-vortex interactions resulting from the placement of a blunt body within a restricted channel. Once again, the model was able to accurately capture the dynamics of the flow. The location of triple point was monitored and matched well with literature.

In the future, the method presented can be directly extended to other conservative PonD schemes (such as DUGKS, DBM, etc) which will allow it to be extended to higher speeds. 
We will also focus on implementing moving boundaries at supersonic and hypersonic speeds. Finally we shall also carry out a thorough analysis of various boundary conditions under the PonD framework. 

\begin{acknowledgments}
     This work was supported by European Research Council (ERC) Advanced Grant No. 834763-PonD. Computational resources at the Swiss National SuperComputing Center CSCS were provided under grant No. s1222.	
\end{acknowledgments}

\section*{Author Declarations}
\subsection*{Conflict of interest}
The authors have no conflicts to disclose.
 
\bibliography{ref}%

\end{document}